\documentclass[aps,prl,superscriptaddress,reprint]{revtex4-1}
\usepackage{amsmath}
\usepackage{amssymb}
\usepackage{graphicx}
\usepackage[colorlinks,urlcolor=blue]{hyperref}
\usepackage{url}
\usepackage{color,xcolor}
\usepackage{ulem}
\usepackage{float}
\usepackage{epstopdf}

\emergencystretch=\maxdimen
\begin{document}

\title{Electronic structure, magnetic properties and pairing tendencies of the copper-based honeycomb lattice Na$_2$Cu$_2$TeO$_6$}
\author{Ling-Fang Lin}
\affiliation{Department of Physics and Astronomy, University of Tennessee, Knoxville, TN 37996, USA}
\author{Rahul Soni}
\affiliation{Department of Physics and Astronomy, University of Tennessee, Knoxville, TN 37996, USA}
\affiliation{Materials Science and Technology Division, Oak Ridge National Laboratory, Oak Ridge, TN 37831, USA}
\author{Yang Zhang}
\affiliation{Department of Physics and Astronomy, University of Tennessee, Knoxville, TN 37996, USA}
\author{Shang Gao}
\affiliation{Materials Science and Technology Division, Oak Ridge National Laboratory, Oak Ridge, TN 37831, USA}
\affiliation{Neutron Scattering Division, Oak Ridge National Laboratory, Oak Ridge, TN 37831, USA}
\author{Adriana Moreo}
\affiliation{Department of Physics and Astronomy, University of Tennessee, Knoxville, TN 37996, USA}
\affiliation{Materials Science and Technology Division, Oak Ridge National Laboratory, Oak Ridge, TN 37831, USA}
\author{Gonzalo Alvarez}
\affiliation{Computational Sciences \& Engineering Division, Oak Ridge National Laboratory, Oak Ridge, TN 37831, USA}
\author{Andrew D. Christianson}
\affiliation{Materials Science and Technology Division, Oak Ridge National Laboratory, Oak Ridge, TN 37831, USA}
\author{Matthew B. Stone}
\affiliation{Neutron Scattering Division, Oak Ridge National Laboratory, Oak Ridge, TN 37831, USA}
\author{Elbio Dagotto}
\affiliation{Department of Physics and Astronomy, University of Tennessee, Knoxville, TN 37996, USA}
\affiliation{Materials Science and Technology Division, Oak Ridge National Laboratory, Oak Ridge, TN 37831, USA}

\begin{abstract}
Spin-$1/2$ chains with alternating antiferromagnetic and ferromagnetic couplings have attracted considerable interest due to the topological character of their spin excitations. Here, using density functional theory and density matrix renormalization group methods, we have systematically studied the dimerized chain system Na$_2$Cu$_2$TeO$_6$. Near the Fermi level, the dominant states are mainly contributed by the Cu $3d_{x^2-y^2}$ orbitals highly hybridized with the O $2p$ orbitals in the nonmagnetic phase, leading to an ``effective'' single-orbital low-energy model. Furthermore, the bandwidth of the Cu $3d_{x^2-y^2}$ states is small ($\sim 0.8$ eV), suggesting that electronic correlations will strongly affect this system. By introducing such electronic correlations, we found this system is a Mott insulator. Moreover, by calculating the magnetic exchange interactions ($J_1$, $J_2$ and $J_3$), we explained the size and sign of the exchange interactions in Na$_2$Cu$_2$TeO$_6$, in agreement with neutron experiments. Based on the Wannier functions from first-principles calculations, we obtained the relevant hopping amplitudes and an ``effective'' $d_{x^2-y^2}$ Wannier function combining O $2p$ states with Cu states. A strong Cu-O-O-Cu super-super-exchange plays the main role for the largest antiferromagnetic exchange coupling, because of the direct overlap of the ``effective'' Wannier functions (combination of Cu $3d_{x^2-y^2}$ and O $2p$ states) along the long-distanced Cu-Cu sites ($J_1$ path). Moreover, the exchange interaction along the $J_2$ path is FM because the Cu-O-Cu angle is closed to $90 ^{\circ}$, where a pair of orthogonal O 2p orbitals with parallel spins are involved in the virtual electron hopping. In addition, we constructed a single-orbital Hubbard model for this dimerized chain system, where the quantum fluctuations are taken into account. Both AFM and FM coupling ($\uparrow$-$\downarrow$-$\downarrow$-$\uparrow$) along the chain were found in our DMRG and Lanczos calculations, in agreement with DFT and neutron results. We also calculated the hole pairing binding energy $\Delta E$ which becomes negative at Hubbard $U \sim 11$ eV, indicating incipient pairing tendencies. Finally, we also looked at various cases of hole doping that always exhibit tight pairs. Thus, we believe our results for Na$_2$Cu$_2$TeO$_6$ could provide guidance to experimentalists and theorists working on this dimerized chain system, such as short-range magnetic coupling, doping effects, and possible pairing tendencies.
\end{abstract}

\maketitle
\section{I. Introduction}
One-dimensional (1D) systems continue to attract considerable interest due to their strong quantum fluctuations, as well as their intertwined charge, spin, orbital, and lattice degrees of freedom~\cite{bertini2021finite,Dagotto:Rmp94,grioni2008recent,Dagotto:Rmp,lin2021origin}. This leads to unusual physical properties, such as superconductivity in copper or iron ladders~\cite{Dagotto:sci96,Dagotto:Rpp,uehara1996superconductivity,Takahashi:Nm,Ying:prb17,Zhang:prb17,Zhang:prb18}, magnetic block states in iron ladders~\cite{zhang2019magnetic,zhang2020iron},
orbital-selective Mott phases in 1D chains and ladders~\cite{rincon2014exotic,zhang2021magnetic,lin2022prediction,herbrych2020block,pandey2020prediction,herbrych2020block1,pandey2021intertwined,sroda2021quantum}, excitonic magnets in multiorbital models on chains~\cite{kaushal2020bcs,kaushal2021magnetization} ferroeletricity in WO$X_4$ ($X$ = halogen element)~\cite{lin2019quasi}, charge density waves in Ta-chains~\cite{gooth2019axionic,zhang2020first}, superconductivity in doped Haldane chains~\cite{patel2020emergence}, edge Majorana states in proximity of superconductivity~\cite{herbrych2021interaction}, orbital order in ruthenates~\cite{hotta2001prediction}, ferromagnetism and phase separation in multiorbital $t-J$ model chains~\cite{riera1997phase}, and exotic orbital and magnetic properties in van der Waals chains~\cite{zhang2022electronic}.

As the simplest systems, spin$-1/2$ chains with alternating antiferromagnetic (AFM) and ferromagnetic (FM) couplings display interesting quantum magnetism and gapped excitations~\cite{hida1992ground,hida1994excitation}. These systems usually do not exhibit long-range order at 0~K, where the two AFM spins form a spin dimer, leading to a spin-singlet ($(|{\uparrow \downarrow}\rangle -|{\downarrow \uparrow}\rangle)/\sqrt{2}$) ground state ~\cite{affleck1989quantum,hida1992ground,uhrig1996magnetic}. Furthermore, other unusual properties are local singlet-triplet (triplon) excitations~\cite{haldane1983nonlinear,affleck1987rigorous}, a hidden string order protected by Z$_2$ $\times$ Z$_2$ global rotations; symmetry~\cite{hida1992crossover,kohmoto1992hidden}, and symmetry-protected topological states~\cite{pollmann2010entanglement}. The resource ground state for measurement-based quantum computation~\cite{miyake2010quantum} is also proposed in the AFM-FM chain systems.

\begin{figure*}
\centering
\includegraphics[width=0.98\textwidth]{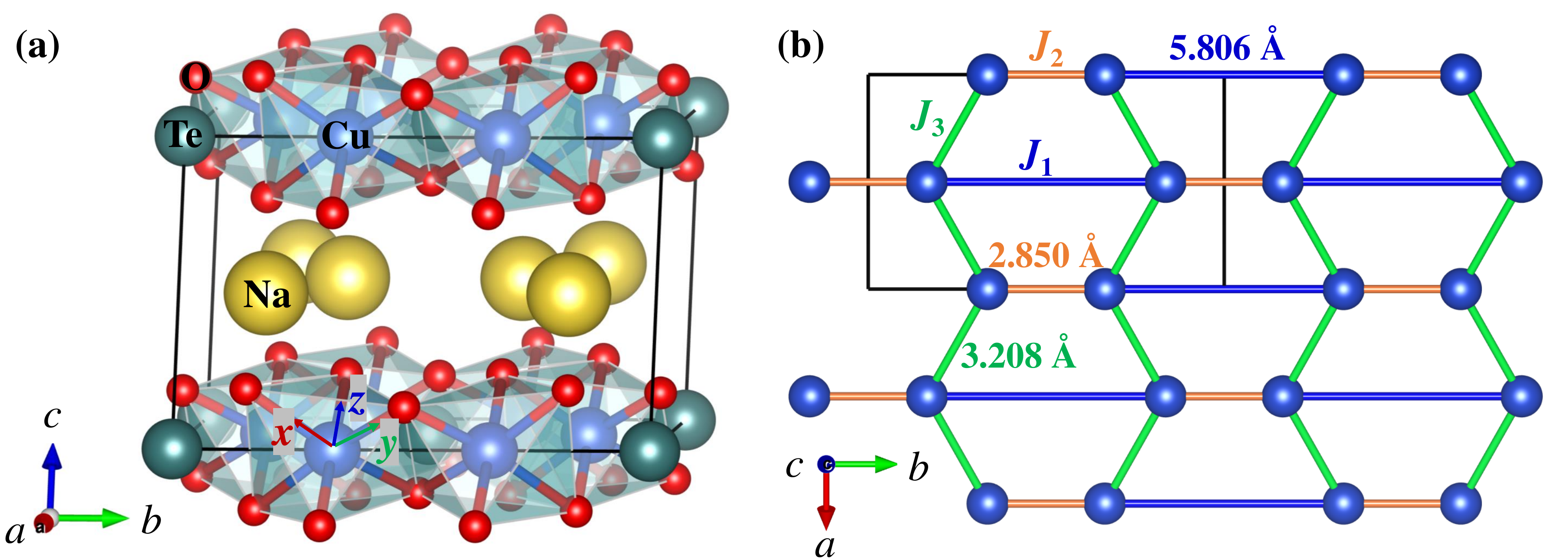}
\caption{Schematic crystal structure of Na$_2$Cu$_2$TeO$_6$: in yellow Na; in  blue Cu; in dark cyan Te; in red O. (a) Conventional cell of the bulk structure. (b) Sketch of the Cu single-layer lattice. Different magnetic exchange couplings are shown in different colors. Note that the local \{$x$, $y$, $z$\} axis are (almost) along the Cu-O bond directions, as marked in Fig.~\ref{structure}(a), leading to $d_{x^2-y^2}$ orbitals spanning over the CuO$_4$ plane.}
\label{structure}
\end{figure*}

However, spin$-1/2$ chains usually display standard staggered AFM couplings due to superexchange Hubbard interactions. To the best of our knowledge, only a few of alternating AFM-FM chains are experimentally realized based on neutron scattering results, including CuNb$_2$O$_6$~\cite{kodama1999neutron}, DMACuCl$_3$~\cite{stone2007quantum}, Na$_3$Cu$_2$SbO$_6$~\cite{miura2008magnetic}, and BaCu$_2$V$_2$O$_8$~\cite{klyushina2018hamiltonian}. Recently, the distorted honeycomb lattice compound, Na$_2$Cu$_2$TeO$_6$, was proposed to be a FM-AFM dimerized chain system~\cite{xu2005synthesis,miura2006spin,derakhshan2007electronic,koo2008analysis,schmitt2014microscopic,gao2020weakly,shangguan2021evidence}. As shown in Fig.~\ref{structure}, Na$_2$Cu$_2$TeO$_6$ has a monoclinic structure with the space group $C2/m$ (No. 12), stacking by alternating Cu$_2$TeO$_6$ and Na layers along the $c$-axis. In each Cu$_2$TeO$_6$ layer, the Cu ions form a distorted honeycomb lattice that is composed of edge-sharing CuO$_6$ octahedra as displayed in Fig.~\ref{structure}(b). In this system, the valence of Cu is $2+$, leading to an effective $S = 1/2$ spin of Cu (corresponding to the $d^9$ electronic configuration). Previous magnetic susceptibility measurements of a powder sample revealed a spin gap $\Delta \sim 127$ K in this system~\cite{xu2005synthesis}, considered to be related to the strong AFM couplings $J_1$~\cite{xu2005synthesis,derakhshan2007electronic,schmitt2014microscopic}. Very recently, a singlet-triplet excitation was reported in Na$_2$Cu$_2$TeO$_6$ single crystals by inelastic neutron scattering experiments~\cite{gao2020weakly}. Based on neutron experiments~\cite{gao2020weakly}, $J_1$ ($\sim 22.78$ meV) is larger than $J_2$ ($\sim -8.73$ meV) although the length between two Cu sites along the $J_1$ path ($5.806$~\AA) is much longer than the one along the $J_2$ path ($2.850$~\AA) [see Fig.~\ref{structure}(b)]. In the dimerized chain direction, the long-distanced Cu-Cu sites form AFM spin-dimers, but not the short-distanced Cu-Cu sites, indicating that O sites must be playing a key role. Furthermore, the interchain coupling $J_3$ is considered much smaller than $J_1$ and $J_2$.

To better understand this interesting system, we have systematically studied the dimerized chain Na$_2$Cu$_2$TeO$_6$ by using first-principles DFT and also DMRG and Lanczos calculations. First, our DFT calculations found that the states near the Fermi level are mainly contributed by Cu $3d$ states with a small bandwidth, which are highly hybridized with O $2p$ orbitals in a nonmagnetic (NM) state, leading to an ``effective'' single-orbital low-energy model. Due to the dimerization in the antibonding $\sigma$ combination of Cu $3d_{x^2-y^2}$ and O $2p$ states, a small gap opens in the band structures of the NM state. In addition, by introducing the electronic correlations, we found this system is a Mott-insulator with a large gap. By mapping the DFT energies to the Heisenberg model, we obtained AFM couplings $J_1$ and $J_3$ while $J_2$ is FM, in agreement with the previously mentioned results. Based on Wannier functions from first-principles calculations, we obtained the relevant hopping amplitudes and an ``effective'' $d_{x^2-y^2}$ Wannier function combined with O $2p$ states. In this case, the AFM spin-dimer for the long-distanced Cu-Cu sites arises from the direct overlap with the ``effective'' Wannier functions (combination of Cu $3d_{x^2-y^2}$ and O $2p$ states), indicating that the strong Cu-O-O-Cu super-super-exchange plays the most important role for the largest magnetic coupling. Furthermore, the Cu-O-Cu angle is close to $90 ^{\circ}$, which leads to the FM character of $J_2$, because a pair of orthogonal O 2p orbitals with parallel spins are involved in the virtual electron hopping.

In addition, we constructed a single-orbital Hubbard model for the dimerized chain, where the quantum fluctuations are taken into account. The block AFM-FM state ($\uparrow$-$\downarrow$-$\downarrow$-$\uparrow$) along the chain was found in our DMRG calculations, in agreement with DFT and neutron results. Considering that superconductivity was widely reported in doped Cu-based compounds with $d^9$ configuration, we also studied the hole-doping in Na$_2$Cu$_2$TeO$_6$. We calculated the binding energy $\Delta E$ and found it becomes negative for Hubbard $U \sim 11$ eV, indicating a possible pairing tendency. However, we believe the pairs are too small to sustain a robust superconductor. Furthermore, we also studied different hole-doping cases, with similar conclusions. Thus, we believe that our results for Na$_2$Cu$_2$TeO$_6$ could arouse the interest to experimentalists and theorists working on this dimerized chain system, such as short-range magnetic coupling, doping effect, and possible pairing tendencies.

\section{II. DFT method}

In the present study, first-principles calculations, using the projector augmented wave (PAW) method, were employed based on DFT, as implemented in the Vienna {\it ab initio} Simulation Package (VASP) code~\cite{Kresse:Prb96,Kresse:Prb99,Blochl:Prb2}. Electronic correlations were considered by using the generalized gradient approximation (GGA) and the revised Perdew-Burke-Ernzerhof (PBEsol) function~\cite{Perdew:Prl,Perdew:Prl08}. The plane-wave cutoff energy was set as $550$~eV. Furthermore, the $k$-point mesh adopted was ${6}\times{4}\times{6}$ for the conventional cell of Na$_2$Cu$_2$TeO$_6$. Note that this $k$-point mesh was tested explicitly to verify that it already leads to converged energies. For the magnetic calculations, on-site Coulomb interactions were considered by using the local spin density approximation (LSDA) plus $U$ with the Liechtenstein formulation for the double-counting term~\cite{liechtenstein1995density}. In addition to the standard DFT calculation discussed thus far, the maximally localized Wannier functions (MLWFs) method was employed using the WANNIER90 code~\cite{marzari1997maximally,mostofi2008wannier90} with the functions centered at the Cu's 3$d_{x^2-y^2}$. All the crystal structures were visualized with the VESTA code~\cite{momma2011vesta}.

\section{III. DFT results}
\subsection{A. Electronic properties}

\begin{figure}
\centering
\includegraphics[width=0.48\textwidth]{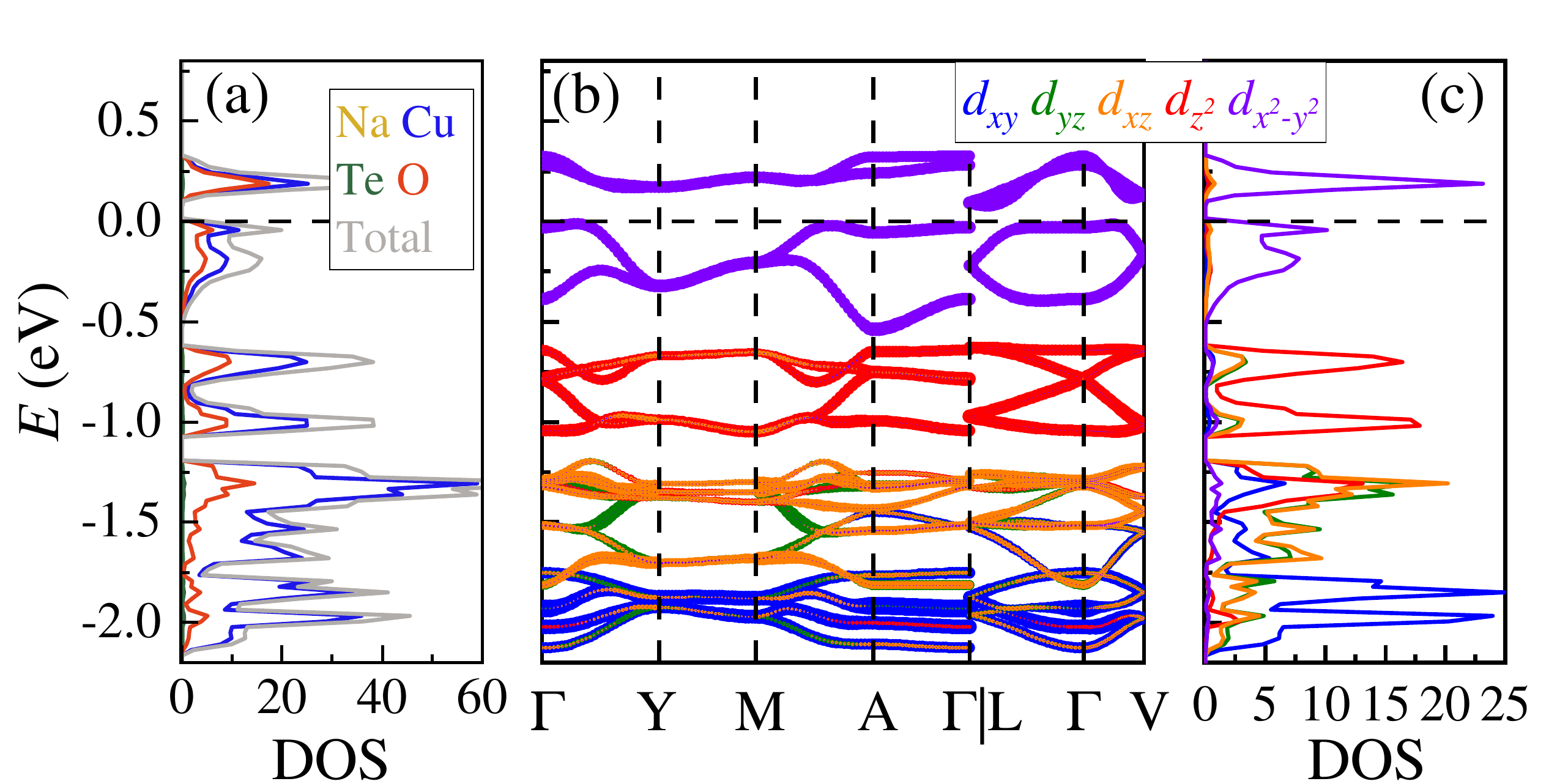}
\caption{(a) DOS near the Fermi level of Na$_2$Cu$_2$TeO$_6$ for the NM phase (in gray Total; in yellow Na; in blue Cu; in dark cyan Te; in red O). (b) Projected band structures and (c) DOS of the NM phase for Na$_2$Cu$_2$TeO$_6$, respectively. Note that the local \{$x$, $y$, $z$\} axes of projected orbitals are marked in Fig.~\ref{structure}. The weight of each Cu orbital is represented by the size of the (barely visible) circles for the projected band structures. The coordinates of the high-symmetry points in the bulk Brillouin zone (BZ) are $\Gamma$ = (0, 0, 0), Y = (0.5,0.5, 0), M = (0.5, 0.5, 0.5), A = (0, 0, 0.5), L = (0, 0.5, 0.5), and V = (0, 0.5, 0) in units of reciprocal basis vectors.}
\label{dosband_nm}
\end{figure}

First, let us discuss the electronic structures for the NM phase of Na$_2$Cu$_2$TeO$_6$.  As shown in Fig.~\ref{dosband_nm}(a), the states near the Fermi level are mainly contributed by the Cu $3d$ orbitals, highly hybridized with the O $2p$ orbitals. Na$_2$Cu$_2$TeO$_6$ turns out to be a charge-transfer system, similar to the cuprate superconductors~\cite{Dagotto:Rmp94,Zhang2020Similarities}. Furthermore, the calculated density of state (DOS) indicates a small gap $\sim 0.09$ eV for Na$_2$Cu$_2$TeO$_6$. This small gap is caused by the dimerization of the antibonding $\sigma$ combination of Cu $3d_{x^2-y^2}$ and O $2p$ states in the distorted honeycomb lattice structure.

Next, to better understand the contribution of Cu $3d$ orbitals, we also calculated the orbital-resolved band structure and DOS. Figures~\ref{dosband_nm}(b) and (c) show that the $d_{x^2-y^2}$ band of Cu's $3d$ is located near the Fermi level (range $-0.5$ eV to $0.3$ eV), while other Cu's $3d$ orbtials ($d_{3z^2-r^2}$, $d_{xz}$, $d_{yz}$, and $d_{xy}$) are fully occupied and at lower energies. In this case, the physical properties of this system are mainly contributed by the $d_{x^2-y^2}$ orbital, i.e. by considering the Cu $3d^9$ configuration in Na$_2$Cu$_2$TeO$_6$. Moreover, the bandwidth $W$ of $d_{x^2-y^2}$ is small ($\sim$ 0.8~eV), leading to a strong electronic correlation effect ($U/W$) in this system. Hence, by introducing the Hubbard $U$, this system should be a Mott-insulator due to the half-filling of the $d_{x^2-y^2}$ orbital of Na$_2$Cu$_2$TeO$_6$, as discussed in the following section.

\begin{figure}
\centering
\includegraphics[width=0.48\textwidth]{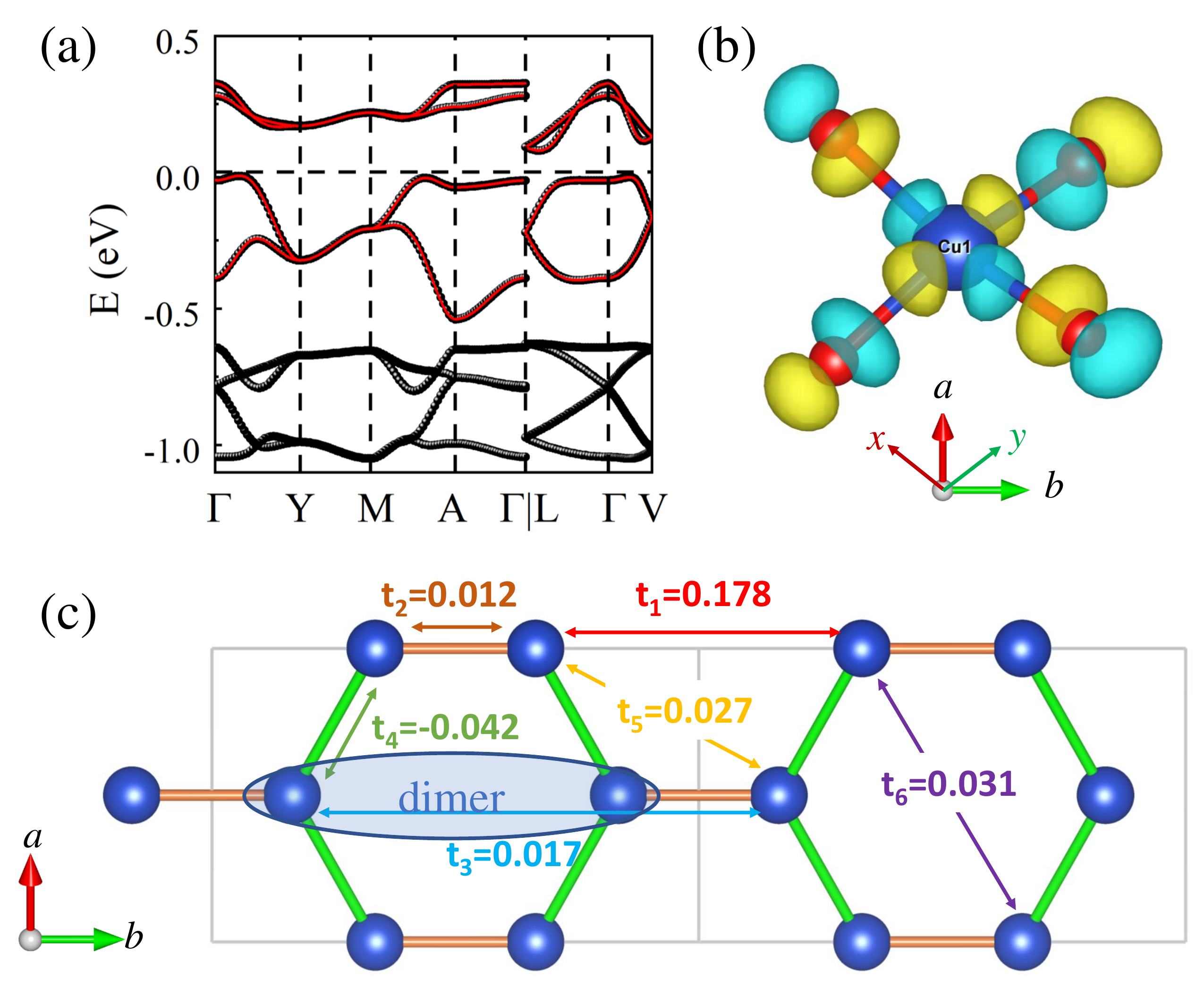}
\caption{(a) DFT (in black) and Wannier bands (in red) of the conventional cell of Na$_2$Cu$_2$TeO$_6$. The Fermi level is shown with dashed horizontal lines. (b) The ``effective'' single orbital of the Wannier function for one site; the isosurface is set to 2. Other Wannier functions on other sites have similar properties, not shown here. (c) The relevant hoppings in the intra $ab$-plane (shown only $|t|>0.01$, in units of eV) based on MLWFs. Note: the inter-layer hoppings are quite small ($\sim 0.014$ eV), and are not shown here.}
\label{hoppings}
\end{figure}

According to the crystal-splitting analysis and electronic structures discussed above, the $d_{x^2-y^2}$ orbital, located near the Fermi level (range $\sim -0.5$ to $\sim 0.3$ eV), determines the physical properties of this system, leading to a single-band low-energy model. To better understand this low-energy model, we constructed one-orbital Wannier functions based on the MLWFs method~\cite{marzari1997maximally,mostofi2008wannier90}, involving a single $d_{x^2-y^2}$ orbital of Cu's $3d$ in the NM phase. Figure~\ref{hoppings}(a) indicates that the single-orbital Wannier band fits very well with the DFT bands. Furthermore, we also plot the ``effective'' single orbital Wannier function for one Cu site, as shown in Fig.~\ref{hoppings} (b). It clearly shows an antibonding combination of $3d_{x^2-y^2}$ and O $2p$ $\sigma$ states. As a result, this ``effective'' single orbital already considers the contribution of O $2p$ states.

Based on the information calculated from MLWFs, the main hoppings between different Cu-Cu sites are obtained as displayed in Fig.~\ref{hoppings}(c), where other small hoppings and inter-layer hoppings are excluded for simplicity. Remarkably, in this system the largest hopping is $t_1 = 0.178$ eV instead of $t_2$ (the short-distanced Cu-Cu sites), indicating that the Cu-O-O-Cu super-super-exchange interaction plays the key role instead of the direct Cu-Cu magnetic interaction. The largest hopping, involving $t_1$, results from the direct overlap of the ``effective'' single-orbital Wannier functions (combination of Cu $3d_{x^2-y^2}$ and O $2p$ states). This is important for the magnetic spin order, as discussed in the next section. The hopping along the $J_2$ path is significantly smaller ($t_2 = 0.012$ eV) than $t_1$, because this hopping originates from the almost orthogonal Wannier functions. In this case, the system forms spin-dimers for the long-distanced Cu-Cu sites [see Fig.~\ref{hoppings}(c)], but not for the short-distanced Cu-Cu sites. Moreover, the inter-layer hoppings are quite small and can be ignored, leading to weak inter-layer magnetic coupling. This is physically reasonable because the magnetic properties are mainly contributed by the single half-filled Cu $3d_{x^2-y^2}$ orbital lying in the $xy$ plane. Due to its layered crystal structure, the overlap between interlayer Cu $3d_{x^2-y^2}$ orbitals are expected to be small.

\subsection{B. Magnetic properties}
\begin{figure}
\centering
\includegraphics[width=0.48\textwidth]{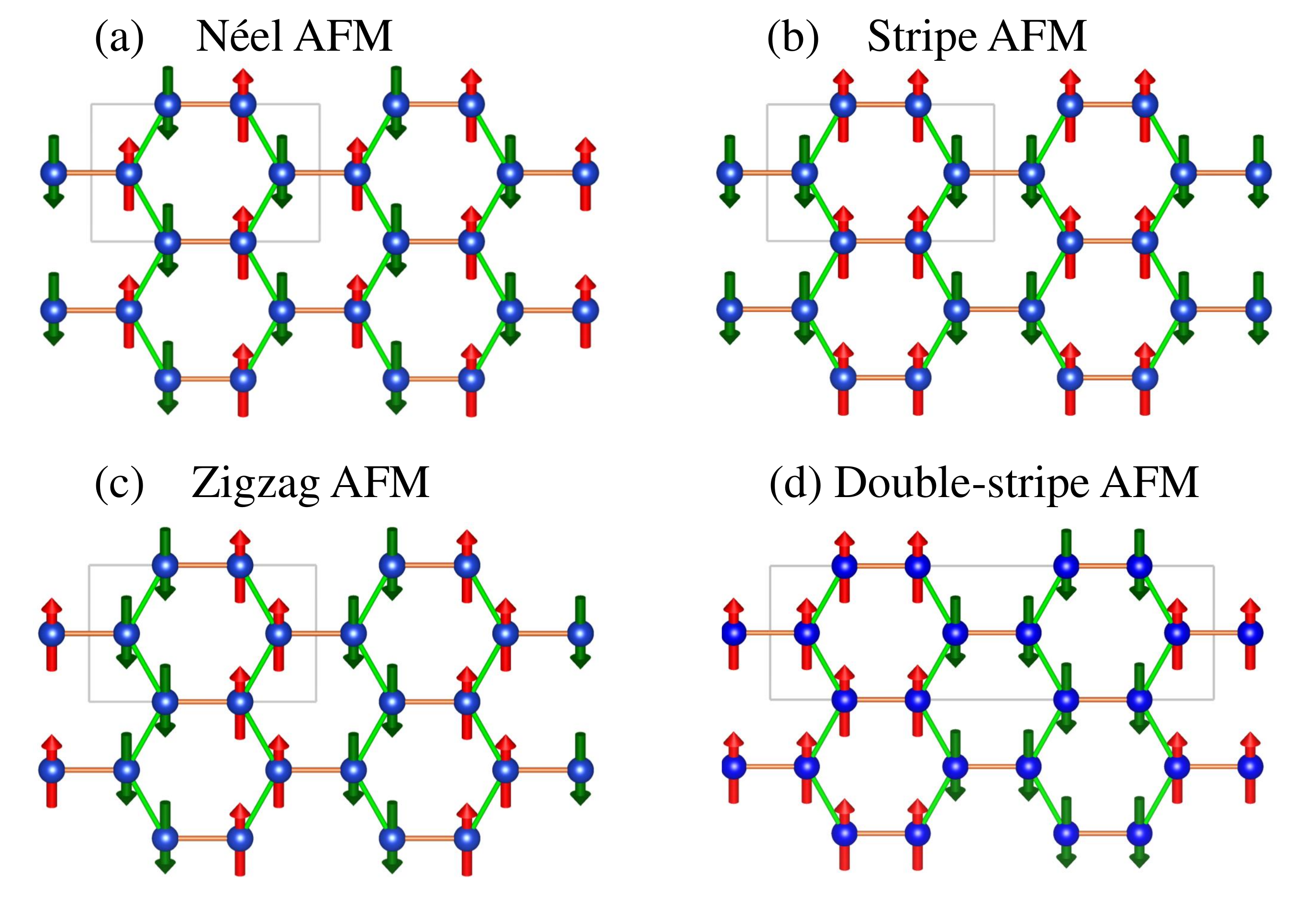}
\caption{Sketch of four possible AFM patterns in the plane studied here. Spin up and down are indicated by red and green arrows, respectively.}
\label{magnetism}
\end{figure}

To better understand the in-plane magnetic properties of Na$_2$Cu$_2$TeO$_6$, we also studied several magnetic configurations in plane, includings FM, N\'{e}el AFM (N-AFM), Stripe AFM (S-AFM), Zigzag AFM (Z-AFM), and Double-stripe AFM (D-AFM) states, as shown in Fig.~\ref{magnetism}. In addition, according to previous experimental results~\cite{gao2020weakly} and hopping analysis, the inter-layer magnetic coupling should be weak and negligible, so that the inter-layer magnetic coupling is considered to be FM in our calculation for simplicity. Here, we introduced the electron correlation by using LSDA plus $U_{\rm eff}$ ($U_{\rm eff} = U-J$) with the Dudarev format on Cu sites~\cite{Dudarev:Prb}.

\begin{figure}
\centering
\includegraphics[width=0.48\textwidth]{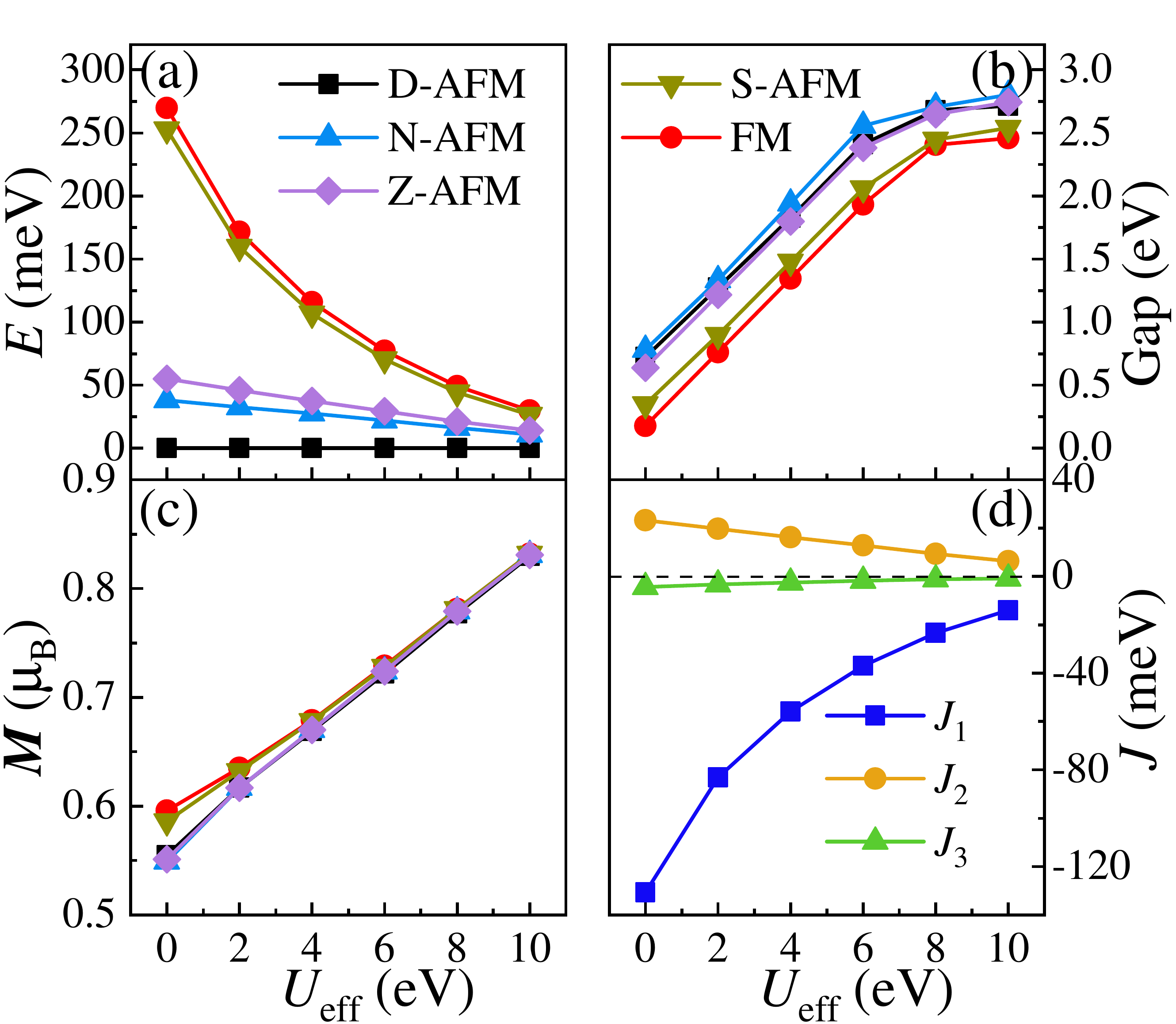}
\caption{DFT results for Na$_2$Cu$_2$TeO$_6$ as a function of $U_{\rm eff}$. (a) Energy (total energy of a supercell, including 8 Cu's) of different spin states. The D-AFM state is taken as the reference. (b) Band gaps of various magnetic orders. (c) Local magnetic moment of Cu calculated within the default Wigner-Seitz sphere. (d) Different magnetic couplings ($J_1$, $J_2$ and $J_3$).}
\label{E_U}
\end{figure}

By using the $1\times2\times1$ supercell of the experimental structure~\cite{gao2020weakly}, we calculated the energies of various magnetic orders as a function of $U_{\rm eff}$ [See Fig.~\ref{E_U}(a)]. Note here the $1\times2\times1$ supercell is the primitive magnetic unit cell to construct the D-AFM state. The D-AFM state always has the lowest energy among all candidate spin configurations, independent of the choice of $U_{\rm eff}$. Furthermore, the band gaps of different magnetic orders are displayed in Fig.~\ref{E_U}(b), where the calculated band gaps are not seriously affected by spin orders. All magnetic ordered states are insulating and the gaps increase with $U_{\rm eff}$, as expected. In addition, the calculated local magnetic moments of Cu of different spin states for different $U_{\rm eff}$ are shown in Fig.~\ref{E_U}(c), in agreement with an $S = 1/2$ with $3d^9$ electronic configuration.

Based on the DFT results, the most important exchange interactions ($J_1$, $J_2$, and $J_3$) are estimated by mapping the calculated DFT energies of different AFM states to the Heisenberg model:
\begin{eqnarray}
\nonumber H&=&-J_1\sum_{<ij>}\textbf{S}_i\cdot\textbf{S}_j-J_2\sum_{[kl]}\textbf{S}_k\cdot\textbf{S}_l\\
&&-J_3\sum_{\{mn\}}\textbf{S}_m\cdot\textbf{S}_n.
\end{eqnarray}
where $J_1$/$J_2$/$J_3$ are the in-plane exchange interactions as marked in Fig.~\ref{structure}(b). Four AFM sates are used to calculate the spin exchange parameters:
\begin{eqnarray}
E_{\rm N-AFM}=E_0+4J_1S^2+4J_2S^2+8J_3S^2,
\end{eqnarray}
\begin{eqnarray}
E_{\rm S-AFM}=E_0-4J_1S^2-4J_2S^2+8J_3S^2,
\end{eqnarray}
\begin{eqnarray}
E_{\rm Z-AFM}=E_0+4J_1S^2+4J_2S^2-8J_3S^2,
\end{eqnarray}
\begin{eqnarray}
E_{\rm D-AFM}=E_0+4J_1S^2-4J_2S^2.
\end{eqnarray}

As shown in Fig.~\ref{E_U}(d), the couplings $J_1$ and $J_3$ are always AFM (negative sign) and $J_2$ is FM (positive sign), depending on $U_{\rm eff}$. In addition, the magnitude of the coupling strength $J_1$ is several times higher than that of the FM coupling $J_2$ and dozens of times higher than that of the AFM coupling $J_3$. In this case, the two strongest spin exchange couplings $J_1$ and $J_2$ lead to alternating AFM-FM chains, which is consistent with
the experimental results~\cite{gao2020weakly}. By changing $U_{\rm eff}$, the ratio $J_2/J_1$ increases in magnitude from $-0.178$ to $-0.446$, while $J_3/J_1$ increases from $0.033$ to $0.057$.
At $U_{\rm eff} = 8$ eV, the calculated strengths of the exchange couplings ($J_2/J_1 = -0.401$ and $J_3/J_1 = 0.052$) are very close to the experimental values ($J_2/J_1 = -0.383$ and $J_3/J_1 = 0.059$).

\begin{figure}
\centering
\includegraphics[width=0.48\textwidth]{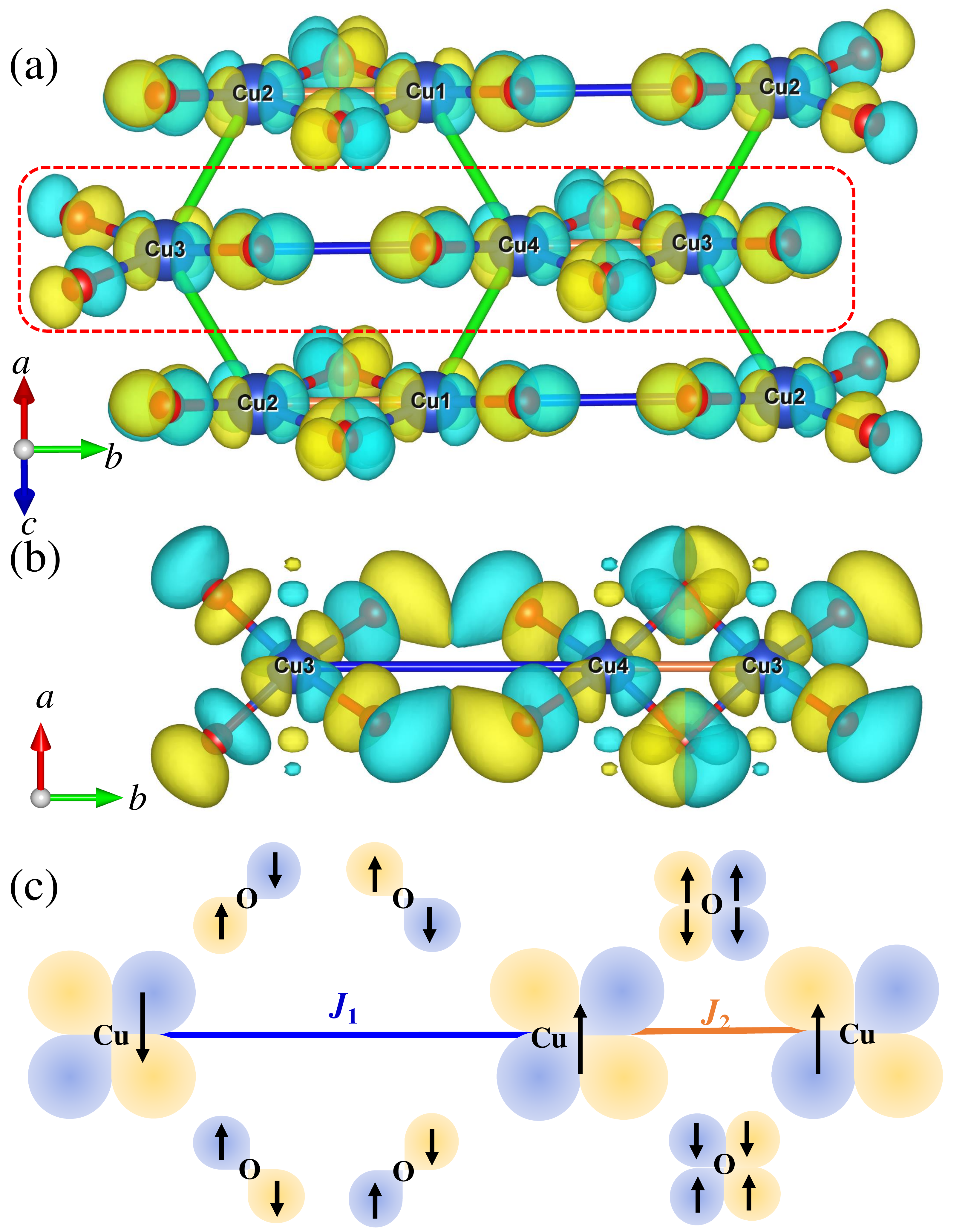}
\caption{(a) Side view of Wannier functions for the Cu $3d_{x^2-y^2}$ orbital for the 2D honeycomb Cu layer of Na$_2$Cu$_2$TeO$_6$. The isosurface is set to be 2. (b) Top view of the Wannier functions of the Cu $3d_{x^2-y^2}$ orbital along the chain direction [red dash rectangle part in (a)]. The isosurface is set to be 0.95. Different colors represent the +/- signs of the Wannier functions. (c) Diagrams for the super-super-exchange and superexchange couplings for different Cu-Cu paths along the chain direction via oxygen $2p$ ligands. For the $J_1$ path, the Cu-O-O-Cu super-super-exchange leads to the AFM alignment of the two Cu ions. For the $J_2$ path, Cu-O-Cu superexchange with a bonding
angle of $90 ^{\circ}$ results in a FM exchange between the nearest-neighbor two ions.}
\label{WF_chain}
\end{figure}

Half-filled systems usually display staggered AFM with the $\uparrow$-$\downarrow$-$\uparrow$-$\downarrow$ spin structure due to the superexchange Hubbard interaction. Although the distance of FM Cu-Cu sites induced by $J_2$ ($\sim 2.850$ \AA) is much shorter than that of the AFM $J_1$ ($\sim 5.806$ \AA), the value of its associated magnetic coupling $J_2$ is several times smaller than that of $J_1$. To understand these DFT and experimental results, we plot the Wannier functions in Fig.~\ref{WF_chain}(a). It clearly shows that the ``effective'' Wannier functions of Cu's $d_{x^2-y^2}$ display strong 1D characteristics, leading to a 1D magnetic chain system. For the interchain $J_3$ path, the superexchange Hubbard interaction leads to an AFM coupling but with little overlap for the Cu-Cu Wannier function along the $J_3$ path. For $J_1$, the magnetic coupling between two Cu sites is along the Cu-O-O-Cu path, leading to a direct overlap of Wannier functions, as displayed in Fig.~\ref{WF_chain}(b). For $J_2$, the magnetic coupling between two Cu sites is the Cu-O-Cu path, resulting in almost orthogonal overlapping Wannier functions [see Fig.~\ref{WF_chain}(b)]. In this case, the $J_1$ path, despite its longer distance, develops a stronger coupling than that over the $J_2$ path, as already explained. Based on this information from the Wannier functions, the signs of the couplings can be understood in Fig.~\ref{WF_chain}(c). For the $J_1$ path, the Cu-O-O-Cu super-super-exchange (two oxygens as the bridge) leads to an AF interaction between two Cu$^{\rm 2+}$ spins. Considering that the Cu-O-Cu angle is close to $90 ^{\circ}$, the interaction becomes FM, because a pair of orthogonal O $2p$ orbitals with parallel spins are involved in the virtual electron hopping. In this case, this system forms weakly coupled alternating AFM-FM $S = 1/2$ chains, instead of a staggered AFM chain.

Considering previous theoretical calculations for other Cu$^{\rm 2+}$ ion materials~\cite{sterling2021effect,pavlenko2007interface,lee2004infinite}, we also calculated the electronic structures of
the D-AFM state of Na$_2$Cu$_2$TeO$_6$ based on LSDA+$U$ with $U_{\rm eff} = 8$ eV. At this $U_{\rm eff} = 8$ eV, the calculated magnetic couplings are $J_1 = 23.39$, $J_2 = -9.38$, and $J_3 = 1.22$ meV, which are in good agreement with the values obtained from neutron experiments ($J_1 = 22.78$, $J_2 = -8.73$, and $J_3 = 1.34$ meV)~\cite{gao2020weakly}. Hence, this value of $U_{\rm eff}$ is reasonable.

\begin{figure}
\centering
\includegraphics[width=0.48\textwidth]{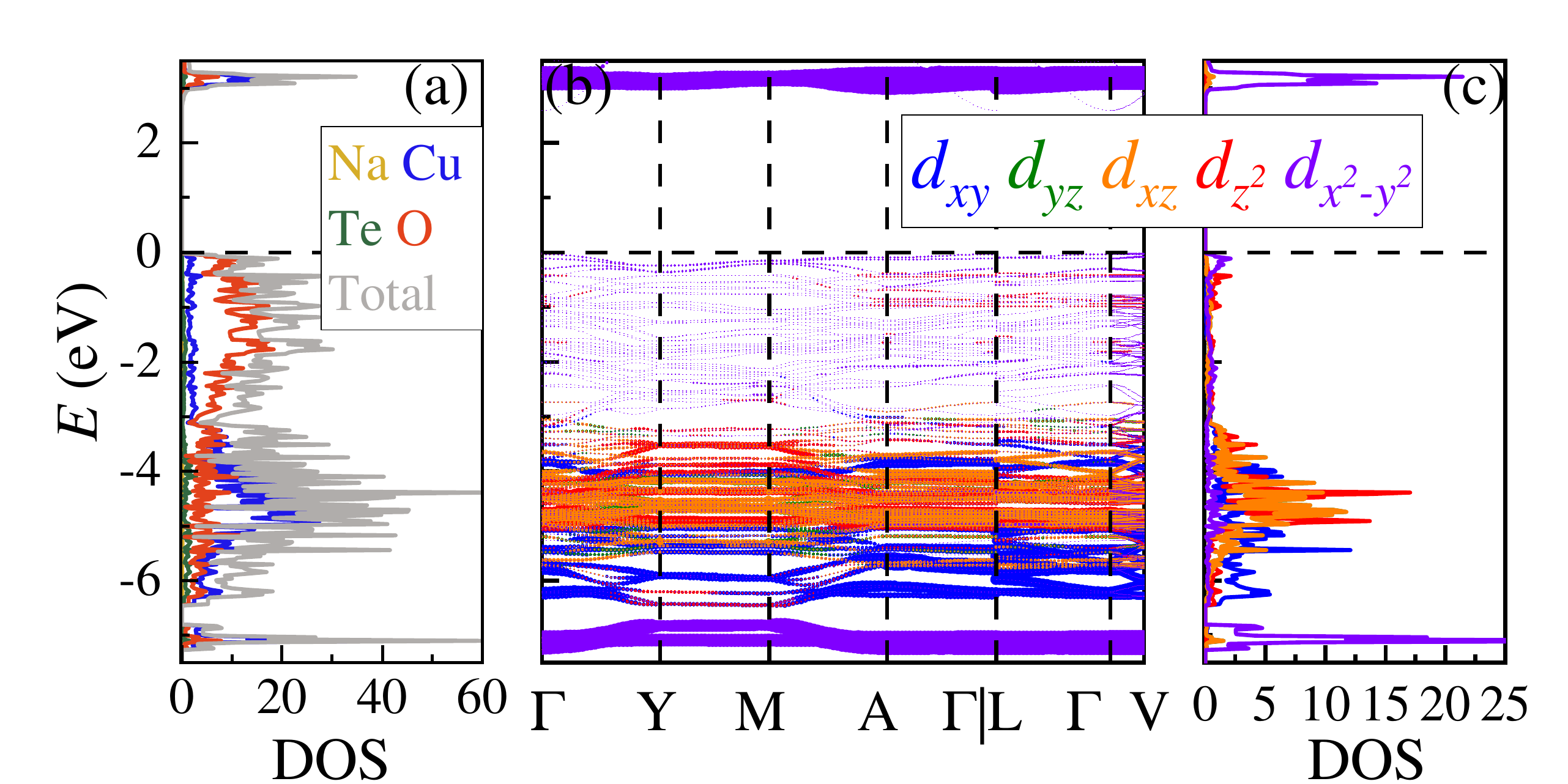}
\caption{(a) DOS near the Fermi level of Na$_2$Cu$_2$TeO$_6$ for the D-AFM phase (Gray: total; yellow: Na; blue: Cu; dark cyan: Te; red: O). (b) Projected band structures and (c) DOS of the D-AFM phase for Na$_2$Cu$_2$TeO$_6$, respectively. Note that the local \{$x$, $y$, $z$\} axes of projected orbitals are marked in Fig.~\ref{structure} (a). The weight of each Cu orbital is represented by the size of the circles for the projected band structures.}
\label{dosband_double}
\end{figure}

As displayed in Fig.~\ref{dosband_double}(a), the Cu $3d$ orbitals shift away from the Fermi level while the O $2p$ states are close to that Fermi level, supporting the charge-transfer picture. Figures~\ref{dosband_double}(b) and (c) indicate that the half-occupied $d_{x^2-y^2}$ orbitals display strong Mott-insulating behavior, while other Cu's $3d$ orbitals are fully-occupied. In this case, this system is locally in a total $S = 1/2$ state, where the magnetism is contributed by the $d_{x^2-y^2}$ state. Furthermore, for the Cu-O-O-Cu $J_1$ path in Na$_2$Cu$_2$TeO$_6$, we also found a small net magnetization at the oxygens of ~0.05 $\mu_{\rm B}$, which originates from the hybridization between atoms and mobility of the electrons, as discussed in Ref.~\cite{lin2021oxygen}.

\section{IV. Single-orbital Hubbard model method}

A wide variety of real materials also have dominant 1D-like physical properties, even without restrictive 1D geometries in their crystal structure. For those systems, interesting phenomena caused by strongly anisotropic electronic structures have been qualitatively unveiled in theory by using simple 1D models, including 1D spin order~\cite{gao2020weakly,lin2021origin}, ferroelectronic distortion ~\cite{lin2019quasi,Lin:Prm}, orbital ordering~\cite{pandey2021origin,lin2021orbital}, nodes in the spin density~\cite{lin2021oxygen}, as well as dimerization ~\cite{zhang2021peierls,zhang2021orbital,zhang2020first}.

To better understand the magnetic coupling for the dimer chain direction, an effective single-orbital Hubbard model was constructed to calculate the real-space spin correlations via the density matrix renormalization group method ~\cite{white1992density,white1993density,schollwock2005density,hallberg2006new}, where we have used the DMRG++ software~\cite{alvarez2009density}. The model studied here includes the kinetic energy and interaction energy terms $H = H_k + H_{int}$:

\begin{eqnarray}
H = \sum_{i,\sigma,{\alpha}}t_{{\alpha}}
(c^{\dagger}_{i\sigma}c^{\phantom\dagger}_{i+{\alpha},\sigma}+H.~c.)+ U\sum_in_{i\uparrow} n_{i\downarrow},
\end{eqnarray}
where the first term represents the hopping of an electron from site $i$ to site $i+{\alpha}$. The number ${\alpha}$ indicates the three different hoppings ($t_1$, $t_2$, and $t_3$), as shown in Fig.~\ref{hoppings}(c). The second term is the standard intraorbital Hubbard repulsion.

Here, we employed a $L=36$-sites chain with open-boundary conditions (OBC). Furthermore, at least 3000 states were kept and up to 17 finite loop sweeps were performed during our DMRG calculations. We also tested other different sizes, such as $L=16, 24, 40$ sites, and the results are robust. The electronic filling $n = 1$ in the active one orbital is considered. This electronic density (one electron in one orbital) corresponds to the total $S = 1/2$ configuration of the $d^9$ configuration of Cu$^{\rm 2+}$. In the tight-binding term, we only considered three hoppings: $t_1 = 0.178$, $t_2 = 0.012$, and $t_3 = 0.017$ (in eV).

\section{V. DMRG results}
\subsection{A. Magnetic properties}
The distorted honeycomb crystal structure studied here is characterized as a low-dimensional spin system due to strong quantum fluctuations~\cite{xu2005synthesis,miura2006spin,shangguan2021evidence}. Because DFT neglects fluctuations, here we adopted the advanced many-body DMRG method to discuss the quantum magnetic coupling in this $S=1/2$ dimerized chain system. To understand the magnetic coupling along the  dimerized chain, we measured the real-space
spin-spin correlations $\langle{{\bf S}_i \cdot {\bf S}_j}\rangle$.
Here the spin at site $i$ is
\begin{eqnarray}
{\bf S}_i = \frac{1}{2}\sum_\gamma\sum_{\alpha\beta}c^\dagger_{i\gamma\alpha}{\boldsymbol \sigma}_{\alpha\beta}c^{\phantom\dagger}_{i\gamma\beta}\,,
\end{eqnarray}
where ${\boldsymbol \sigma}_{\alpha\beta}$ are the matrix elements of the Pauli matrices.

\begin{figure}
\centering
\includegraphics[width=0.48\textwidth]{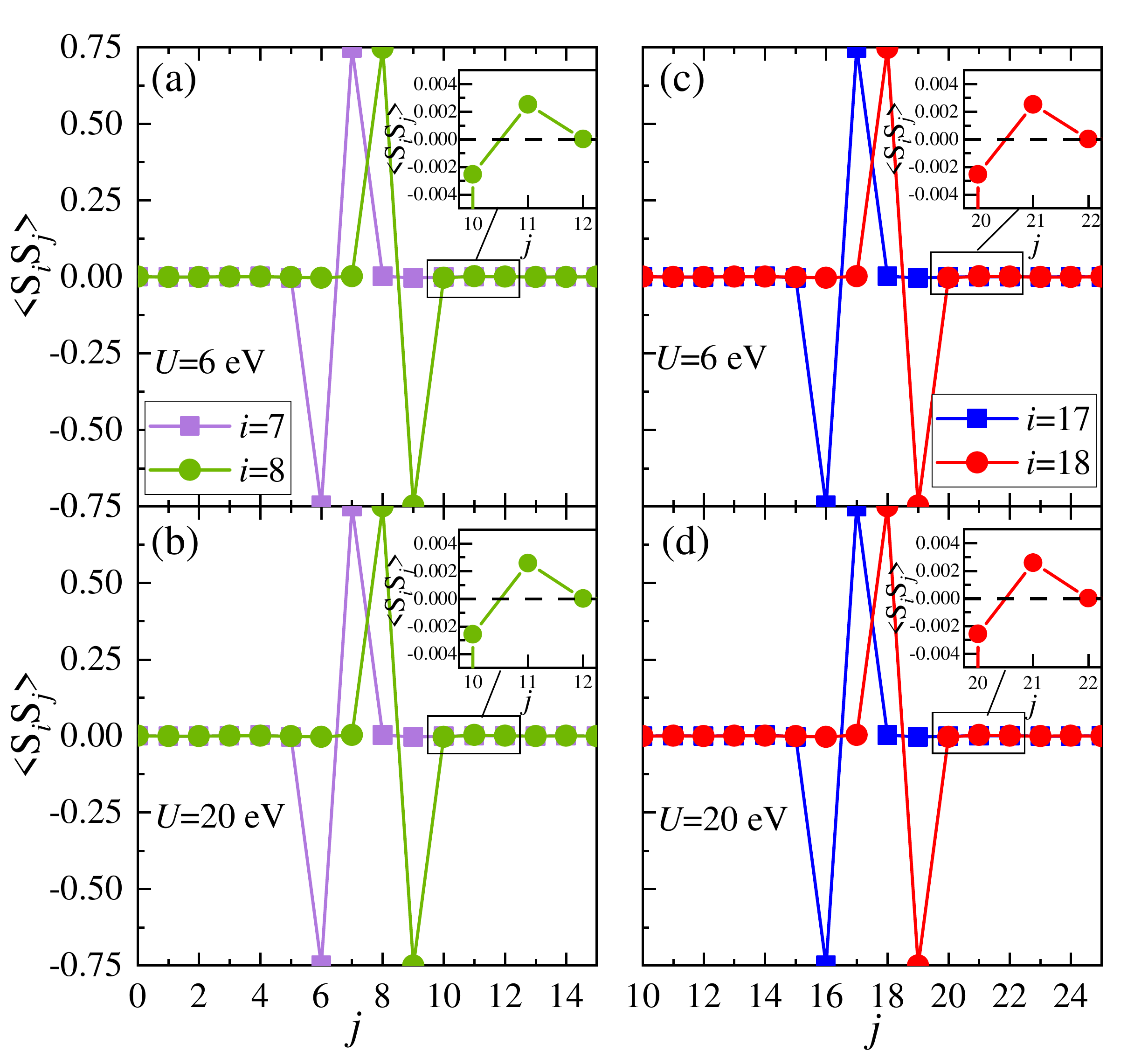}
\caption{(a-b) Spin-spin correlations $\langle{{\bf S}_i \cdot {\bf S}_j}\rangle$ in real space for (a) $U = 6$ eV  and (b) $U = 20$ eV, using $L = 16$. (c-d) Spin-spin correlations $\langle{{\bf S}_i \cdot {\bf S}_j}\rangle$ in real space for (c) $U = 6$ eV  and (d) $U = 20$ eV, using $L = 36$. Insets: the FM coupling between dimers is shown.}
\label{SS_dmrg}
\end{figure}

Figure~\ref{SS_dmrg} shows the spin-spin correlation $\langle{{\bf S}_i \cdot {\bf S}_j}\rangle$ vs. site index for different values of $U$ and length $L$. The distance is $r=\left|{i-j}\right|$, with $i$ and $j$ site indexes. The spin-spin correlation decays very fast with distance $r$, suggesting a long-range disordered phase in this dimerized chain because it is composed of strong dimer spin-singlet states ($(|{\uparrow \downarrow}\rangle -|{\downarrow \uparrow}\rangle)/\sqrt{2}$), nearly decoupled from one another. As shown in the inset of Figs.~\ref{SS_dmrg} (a) and (b), the coupling between dimers is FM coupling, but weak for $L = 16$. Furthermore, we also studied the $L = 36$ case. These results are similar to the results of $L = 16$, indicating
that our conclusions of spin-singlet state, nearly decoupled from one another with weak FM, are robust against changes in $L$.

In the range of $U/W$ studied here, we observed a robust AFM-FM coupling along the chain direction. This AFM-FM coupling chain is reasonable. The magnetic coupling in a dimer should be AFM because the large overlap of Cu-$3d_{x^2-y^2}$ orbitals establishes AFM coupling in a dimer according to the Cu-O-O-Cu super-super-exchange ideas. Between neighboring Cu-Cu dimers, our DMRG calculations predict a short-range coupling, which is FM due to the geometrically nearly perpendicular Cu-O-Cu bond, instead of the direct Cu-Cu AFM interaction, albeit much weaker.

\subsection{B. Binding energy}
Considering that superconductivity was widely reported in doped Cu-based compounds with the $d^9$ electronic configuration~\cite{Dagotto:Rmp94}, we also studied the case of hole doping in Na$_2$Cu$_2$TeO$_6$. To explore possible pairing tendencies, we studied the binding energy of a pair of holes defined as~\cite{Dagotto:Rmp94}:
\begin{eqnarray}
{\Delta}E=E(N-2)+E(N)-2E(N-1),
\end{eqnarray}
where $E(N)$ is the ground-state energy of the undoped case with half-filling for the single-orbital chain model. $E(N-2)$ and $E(N-1)$ are the ground-state energy of the two-hole doped or one-hole doped cases.
Here, $\Delta E$ is negative, indicating pairing tendencies, because the particles minimize their energy by creating a bound state. However, if the holes become two independent particles, this
corresponds to zero binding energy in the bulk limit. In the case where the particles do not bind, this quantity is positive for finite systems and should converge to zero as the size of the cluster increases.

\begin{figure}
\centering
\includegraphics[width=0.48\textwidth]{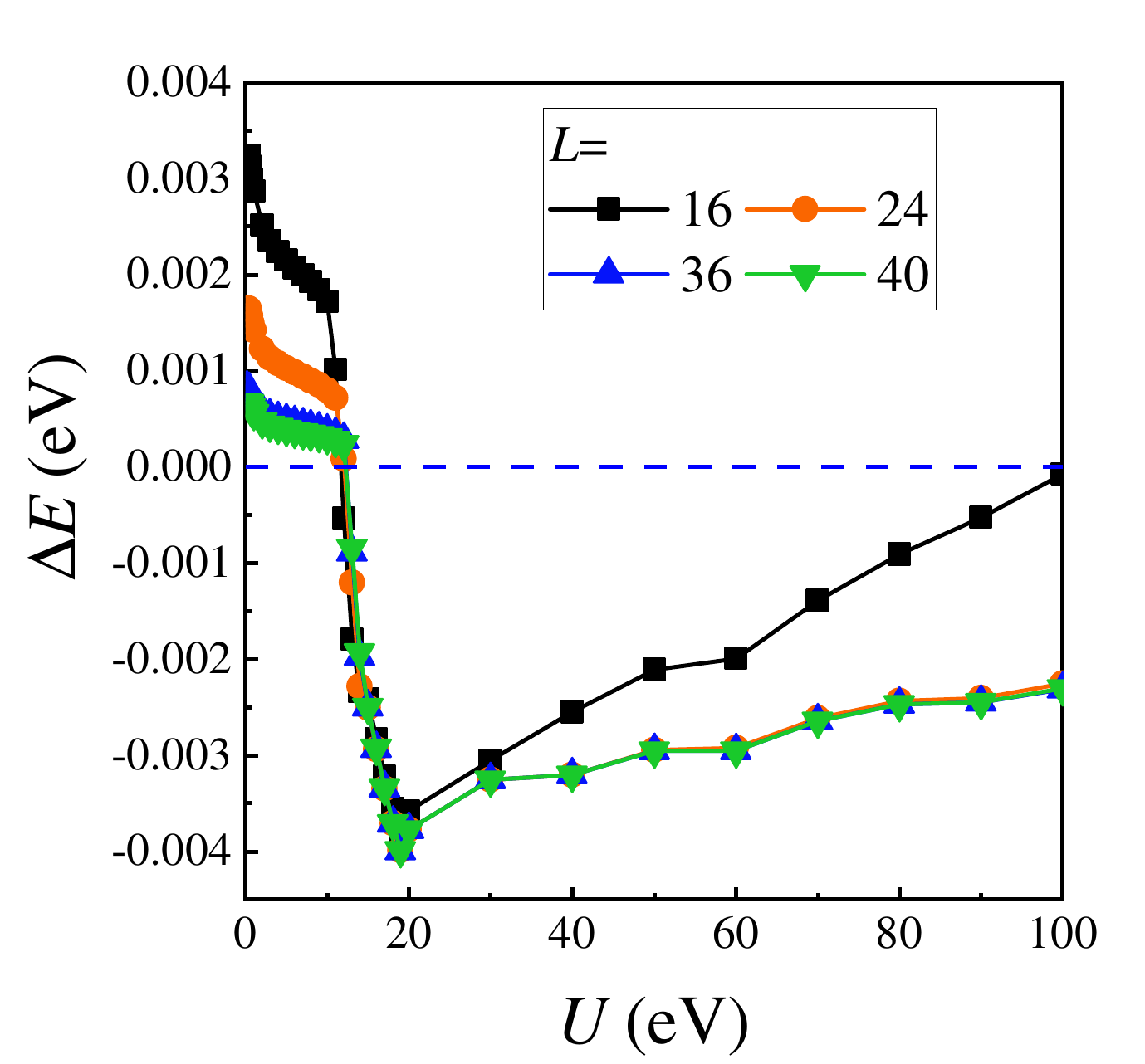}
\caption{Binding energy vs $U$ calculated with DMRG for different chain lengths $L$. In all cases shown, we
observed the possible pairing tendencies in the strong Hubbard $U$ region. The minimum of the binding energy (strongest binding tendency) can be found at $U \sim 20$ eV in all cases.}
\label{BE_dmrg}
\end{figure}

Based on the calculated ground-state energies for the cases $N$, $N - 1$ (one hole), and $N - 2$ (two holes), we obtain the binding energy $\Delta E$ for different chain lengths $L$, as shown in
Fig.~\ref{BE_dmrg}. The results clearly show that the binding energy $\Delta E$ becomes negative in the region Hubbard $U \sim 11$ eV and larger, displaying a broad binding region in Fig.~\ref{BE_dmrg}. In addition, the minimum
of the binding energy $\Delta E$ is found at about $U  \sim 20$ eV. Here, the absolute value of binding energy $|\Delta E|$ is quite small due to the very tiny hopping $t_2$ between singlet dimers.

\begin{figure}
\centering
\includegraphics[width=0.48\textwidth]{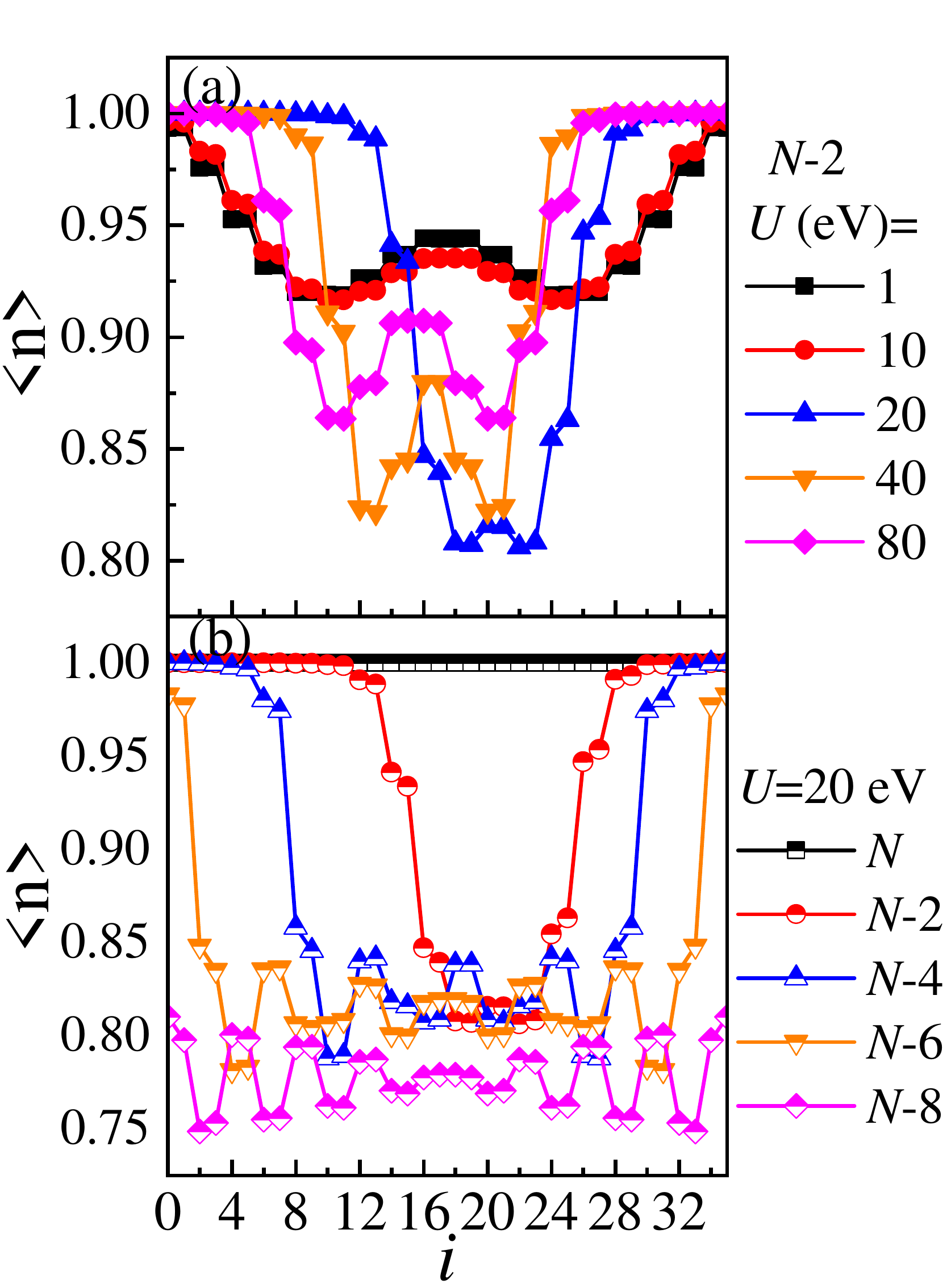}
\caption{(a) Real-space electronic density $n$ of the $N-2$ case for different Hubbard interactions $U$. (b) Real space electronic density $n$ for different hole-doping cases at $U = 20$ eV. Here, we used a chain length $L = 36$.}
\label{CharDen_dmrg}
\end{figure}

To better understand the pairing implication of the negative binding energy obtained from Fig.~\ref{BE_dmrg}, we have also calculated the real-space distribution of charge density in the doped system.

Figure~\ref{CharDen_dmrg} (a) shows the electronic density $n$ of the $N-2$ case for different Hubbard interactions $U$. In the small $U$ region, the electronic density $n$ indicates that the hole density ($1-n$) wants to spread apart. In this case, as a consequence, no pairing in this region $U \leq 10$ eV was found. However, as $U$ increases, the hole density of the pair of holes (the minima) get closer, suggesting that holes prefer to be together, corresponding to the region of negative binding energy.
It is also shown that at $U = 20$~eV the holes are closer than at other values of  $U$. Qualitatively, this kind of results resemble the binding energy because there is more binding at $U = 20$ eV than at other $U$.

In addition, we also studied the real-space electronic density $n$ for different hole-doping cases at $U = 20$ eV [see Fig.~\ref{CharDen_dmrg} (b)]. For $N$ electrons, corresponding to the half-filled orbital, the electronic density is uniform at $n = 1$ for different sites $i$. In the case of $N-2$ electrons, i.e., two holes, these two holes are located near the center of the cluster, in a tight manner compatible with the small pairing.

\section{VI. Lanczos results}
\begin{figure}
\centering
\includegraphics[width=0.48\textwidth]{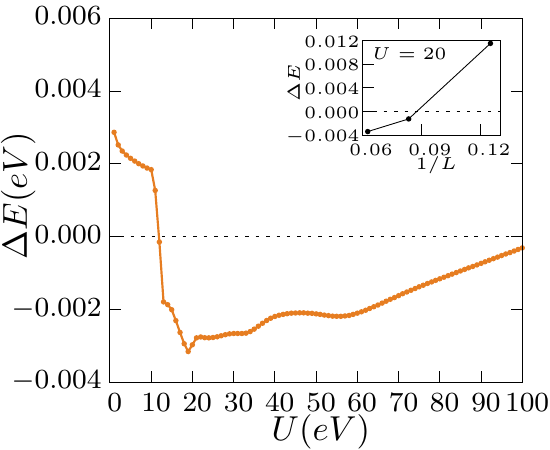}
\caption{Binding energy $\Delta E$ vs $U$ for a 16-site cluster via Lanczos. The inset shows the growth of binding as we increase the system size at $U=20$ (in eV). For the inset plot, system sizes $L=8,12,$ and $16$ were considered.}
\label{Fig: BE vs U 16 site lanczos}
\end{figure}

We also performed Lanczos studies on a 16-site cluster, complementary to our DMRG results. In Fig.~\ref{Fig: BE vs U 16 site lanczos}, we show the binding energy ($\Delta E$) versus the interaction strength $U$. Firstly, consistent with our DMRG results the binding energy behaves quite similarly, with the maximum binding happening at $U\sim 20$ eV. Secondly, the figure in the inset shows the robustness of the binding energy at $U=20$ eV, as we increase the system size (points shown are for $L=8,12$ and $16$ sites). This is an important observation since computationally only small lattice sizes can be studied via Lanczos, and even within this limitation we observe $\Delta E$ becoming more negative as we increase the system sizes. Note that for all these Lanczos results the maximum convergence error is of the order of $10^{-8}$.

\begin{figure}
\centering
\includegraphics[width=0.48\textwidth]{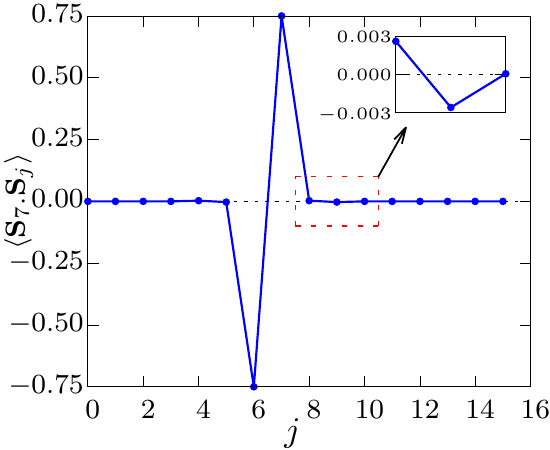}
\caption{Real-space spin-spin correlations with respect to site-$7$ at $U=20$ for a 16-site cluster via Lanczos.}
\label{Fig: S7Sj vs j 16 site lanczos}
\end{figure}

Similar to our DMRG study in the previous section, we have computed the real-space spin-spin correlation for a 16-site chain via Lanczos, see Fig.~\ref{Fig: S7Sj vs j 16 site lanczos}. We observed that our results are in good agreement with DMRG results, providing further confirmation to our study.

\section{VII. Conclusions}
In this publication, we have systematically studied the dimerized chain system Na$_2$Cu$_2$TeO$_6$ by combining first-principles DFT as well as DMRG and Lanczos calculations. Based on the {\it ab initio} DFT calculations for the non-magnetic state, we found the states near the Fermi level are mainly contributed by the Cu $3d$ states highly hybridized with O $2p$ orbitals,  leading to an ``effective'' one-orbital low energy model. Furthermore, we also observed a small gap which is caused by the dimerization of the antibonding $\sigma$ combination of Cu $3d_{x^2-y^2}$ and O $2p$ states in this system. In addition, the bandwidth of the Cu $3d_{x^2-y^2}$ states is small ($\sim 0.8$ eV), indicating a strong electronic correlation effect in this system as $U$ increases. As a result, we also found this system is a Mott insulator. Based on different magnetic configurations, we obtained three magnetic exchange interactions ($J_1$, $J_2$ and $J_3$) by mapping the DFT energies to a Heisenberg model. In this case, $J_1$ and $J_3$ are AFM couplings and $J_2$ is FM, in agreement with the experimental results.

Based on the Wannier functions from first-principles calculations, we obtained the relevant hopping amplitudes and an ``effective'' $d_{x^2-y^2}$ Wannier function in combination with O $2p$ states, leading to a spin-singlet formation in an AFM dimer. In this AFM dimer, the strong Cu-O-O-Cu super-super-exchange plays the main role in generating the largest AFM coupling between the long-distanced Cu-Cu sites, due to the direct overlapping of the ``effective'' Wannier functions (combination of Cu $3d_{x^2-y^2}$ and O $2p$ states). Furthermore, the exchange interaction of the $J_2$ path is FM because the Cu-O-Cu  angle is close to $90 ^{\circ}$, i.e. a pair of orthogonal O $2p$ orbitals with parallel spins are involved in the virtual electron hopping.

In addition, we constructed a single-orbital Hubbard model for this dimerized chain system, where the quantum fluctuations are taken into account. The AFM-FM magnetic coupling ($\uparrow$-$\downarrow$-$\downarrow$-$\uparrow$) along the chain was found in our DMRG calculations, in agreement with DFT calculations and neutron scattering results. Considering that superconductivity was widely reported in doped Cu-based compounds with $d^9$ configuration, we also studied hole doping in Na$_2$Cu$_2$TeO$_6$. We calculated the binding energy $\Delta E$ that becomes negative from Hubbard $U \sim 11$ eV, indicating a possible pairing tendency, forming very small size Cooper pairs. Furthermore, we also studied several hole-doping cases, still suggesting that the pairing tendency is robust. Because the hole pairs are so tight, very likely the critical temperature related to this material will be very small. Overall, our results for Na$_2$Cu$_2$TeO$_6$ will provide guidance to experimentalists and theorists working on this dimerized chain system, such as short-range magnetic coupling, doping effects, and possible pairing tendencies.

\section{Acknowledgments}
The work of L.-F.L., R.S., Y.Z., S. G., A.M, A. D. C., M. B. S. and E.D. was supported by the U.S. Department of Energy (DOE), Office of Science, Basic Energy Sciences (BES), Materials Sciences and Engineering Division. G.A. was partially supported by the Scientific Discovery through Advanced Computing (SciDAC) program funded by U.S. DOE, Office of Science, Advanced Scientific Computing Research and BES, Division of Materials Sciences and Engineering.

\bibliographystyle{apsrev4-1}
\bibliography{ref3}
\end{document}